\documentstyle [12pt]{article}
\def\fnote#1#2{\begingroup\def\thefootnote{#1}\footnote{#2}\addtocounter
{footnote}{-1}\endgroup}
\def\inbar{\vrule height1.5ex width.4pt depth0pt}
\def\IB{\relax{\rm I\kern-.18em B}}
\def\IC{\relax\,\hbox{$\inbar\kern-.3em{\rm C}$}}
\def\ID{\relax{\rm I\kern-.18em D}}
\def\IE{\relax{\rm I\kern-.18em E}}
\def\IF{\relax{\rm I\kern-.18em F}}
\def\IG{\relax\,\hbox{$\inbar\kern-.3em{\rm G}$}}
\def\IH{\relax{\rm I\kern-.18em H}}
\def\II{\relax{\rm I\kern-.18em I}}
\def\IK{\relax{\rm I\kern-.18em K}}
\def\IL{\relax{\rm I\kern-.18em L}}
\def\IM{\relax{\rm I\kern-.18em M}}
\def\IN{\relax{\rm I\kern-.18em N}}
\def\IO{\relax\,\hbox{$\inbar\kern-.3em{\rm O}$}}
\def\IP{\relax{\rm I\kern-.18em P}}
\def\IQ{\relax\,\hbox{$\inbar\kern-.3em{\rm Q}$}}
\def\IR{\relax{\rm I\kern-.18em R}}
\def\ZZ{\relax{\sf Z\kern-.4em Z}}
\def\fnote#1#2{\begingroup\def\thefootnote{#1}\footnote{#2}\addtocounter
{footnote}{-1}\endgroup}
\def\beq{\begin{equation}}
\def\eeq{\end{equation}}
\def\bea{\begin{eqnarray}}
\def\eea{\end{eqnarray}}
\def\lleq#1{\label{#1}\eeq}
\let\nn=\nonumber
\def\tabroom{\hbox to0pt{\phantom{\Huge A}\hss}}
\def\notin{\ \hbox{{$\in$}\kern-.51em\hbox{/}}}
\def\a{\alpha}        
     
     \def\si{\sigma}
   \def\th{\theta}

   \def\cM{{\cal M}}
\def\cN{{\cal N}}

\def\tchi{\tilde \chi}

   \def\bz{{\bar z}}

 \def\bPhi{{\bar \Phi}}

\def\bth{{\bar \theta}}

\def\picture #1 by #2 (#3){
  $$\vbox to #2{
    \hrule width #1 height 0pt depth 0pt
    \vfill
    \special{picture #3}}$$\vspace{-0.5cm}}
\def\scaledpicture #1 by #2 (#3 scaled #4){{
 \dimen0=#1 \dimen1=#2
 \divide\dimen0 by 1000 \multiply\dimen0 by #4
 \divide\dimen1 by 1000 \multiply\dimen1 by #4
 \picture \dimen0 by \dimen1 (#3 scaled #4)}}

\renewenvironment{thebibliography}[1]
  { \begin{list}{\arabic{enumi}.}
    {\usecounter{enumi} \setlength{\parsep}{0pt}
     \setlength{\itemsep}{3pt} \settowidth{\labelwidth}{#1.}
     \sloppy
    }}{\end{list}}
\begin{document}
\hfill {
{HD--THEP--92--47}}
\vskip .05truein
\hfill {
{October 1992}}

\vskip .4truein
\parindent=1.5pc

\begin{center}{
   {\large {\bf Noncritical Dimensions for Critical String Theory: \\
                    Life beyond the Calabi--Yau Frontier}
\fnote{\diamond}{Based in part on talks presented at the NATO ARW
                 on Low Dimensional Topology and Quantum Field Theory
                Cambridge, England, 1992, and the International Workshop
                on String Theory, Quantum Gravity and the Unification of
                Fundamental Interactions, Rome, Italy, 1992.}
}\\
\vglue 1cm
{Rolf Schimmrigk}\\ [2ex]
\baselineskip=16pt
{\it Institut f\"ur Theoretische Physik, Universit\"at Heidelberg}\\
{\it Philosophenweg 16, 6900 Heidelberg, FRG}\\
\vglue 3.5cm
{ABSTRACT}}
\end{center}
\vglue 0.3cm
{\rightskip=1.2pc
 \leftskip=1.2pc
\baselineskip=14pt
\noindent
A recently introduced framework for the compactification of
supersymmetric string theory involving noncritical manifolds of complex
dimension $2k+D_{crit}$, $k\geq 1$, is reviewed. These higher dimensional
manifolds are spaces with quantized positive Ricci curvature and
therefore do not, a priori, describe consistent string vacua. It is
nevertheless possible to derive from these manifolds the massless spectra
of critical string groundstates. For a subclass of these noncritical
theories it is also possible to explicitly construct Calabi--Yau manifolds
from the higher dimensional spaces. Thus the new class of theories makes
contact with the standard framework of string compactification. This class
of manifolds is more general than that of Calabi--Yau manifolds because it
contains spaces which correspond to critical string vacua with no K\"ahler
deformations, i.e. no antigenerations, hence providing mirrors of rigid
Calabi--Yau manifolds. The constructions reviewed here lead to new insight
into the relation between exactly solvable models and their mean field
theories on the one hand and Calabi--Yau manifolds on the other, leading,
for instance, to a modification of Gepner's conjecture.
They  also raise fundamental questions about the Kaluza--Klein concept of
string compactification, in particular regarding the r\^{o}le played by
the dimension of the internal theories.
\vglue 0.8cm}
\renewcommand\thepage{}
\vfill \eject
{\bf\noindent 1. Introduction}
\vglue 0.2cm
\baselineskip=18.2pt
\pagenumbering{arabic}

\noindent
String theory remains the only viable candidate for a unified theory of
quantum gravity. One of the attractions of this theory is the fact that
it describes a rather tight framework. A consequence is that there are
severe restrictions on the internal part of the theory which to a large
extent determines the observable low energy physics in four dimensions.
Based on the conventional framework formulated in \cite{chsw} it
is believed that in left--right symmetric compactifications without
torsion
the internal space of the heterotic string is described by a space which
has to be a compact manifold which is
\begin{itemize}
 \item complex,
 \item K\"ahler, and admits a
 \item covariantly constant spinor,
\end{itemize}
 i.e. has vanishing first Chern class, so--called Calabi--Yau manifolds.

Such manifolds are particularly simple, a  fact that is encoded
concisely
in the spectrum of the theory described, in part, by the cohomology
of the
space. Because the space is complex the real cohomology can be
decomposed,
via the Hodge decomposition, into complex cohomology groups. Thus the
Betti numbers $b_i={\rm dim}_{\IR}~ H^i(\cM,\IR)$ can be expressed in
terms of
the Hodge numbers
$h^{(p,q)}={\rm dim}_{\IC}~ H^{p,q}(\cM, \IC)$:
\beq
b_i = \sum_{p+q=i} h^{(p,q)}.
\eeq
Because the manifold is K\"ahler the Hodge numbers are symmetric,
$h^{(p,q)}=h^{(q,p)}$, and because the first Chern class vanishes
it follows
that $h^{(p,0)}=0=h^{(0,p)}$ for  $p=1,2$ and $h^{(3,0)}=1=h^{(0,3)}$.
Hence the cohomology
of the internal space, summarized in the Hodge diamond

\begin{center}
\begin{tabular}{c c c c c c c}
  &  &          &1         &              &   &   \\
  &  &0         &          &0             &   &   \\
  &0 &          &$h^{(1,1)}$ &            &0  &   \\
1 &  &$h^{(2,1)}$ &          &$h^{(2,1)}$ &   &1  \\
  &0 &          &$h^{(1,1)}$ &              &0  &   \\
  &  &0         &          &0             &   &   \\
  &  &          &1         &              &   &   \\
\end{tabular}
\end{center}

\noindent
contains only two independent elements $h^{(1,1)}=h^{(2,2)}$ and
$h^{(2,1)}=h^{(1,2)}$ which parametrize the number of antigenerations
and
generations, respectively, that are observed in low energy physics.

It is also believed that this class of
string vacua features an unexpected symmetry, mirror symmetry, which
has been discovered in the context of Landau--Ginzburg vacua in
\cite{cls}. Independent evidence
for this symmetry has been found in the context of orbifolds of
exactly solvable tensor models by Greene and Plesser \cite{gp1}. The
effect of this symmetry is that for each string vacuum with some
number $h^{(1,1)}$ of antigenerations and some number $h^{(2,1)}$ of
generations there
exists a mirror vacuum for which these number are exchanged: the
spectrum
of the mirror vacuum consists of $h^{(2,1)}$ antigenerations and
$h^{(1,1)}$ generations. Mirror symmetry thus flips the Hodge diamond
along the offdiagonal.

Mirror symmetry is by now well established: beyond the class of exactly
solvable models discussed in \cite{gp1}, in which mirror symmetry is
understood best, lies the much larger class of Landau--Ginzburg theories
constructed explicitly in \cite{cls}
\fnote{1}{The construction of all quasihomogeneous $N=2$
Landau--Ginzburg
          with an isolated singularity was recently completed
          in \cite{ar,maha}.}
which clearly indicates that mirror symmetry
is a property of string theory. That this symmetry is not accidental
in this
wider context has been proven in \cite{ls4} where it was shown that
by a combination
of orbifolding and fractional transformations a mirror construction
can be
established between a priori independent pairs of Landau--Ginzburg
theories with opposite spectrum. Mirror symmetry is at present being
used
as a hypothesis to obtain
results in algebraic geometry and has been shown to be correct in all
computations that have been performed sofar
\cite{pxpl,norwegians,sheldon}.

Mirror symmetry creates a puzzle.
There exist well--known Calabi--Yau vacua which are rigid, i.e. they do
not have string modes corresponding to complex deformations of the
manifold. Since mirror symmetry exchanges complex deformations and
K\"ahler
deformations of a manifold
it would seem that the mirror of a rigid Calabi--Yau manifold cannot be
K\"ahler and hence does not describe a consistent string vacuum. In
fact, it appears, using Zumino's result \cite{zumino} that $N=2$
supersymmetry of a
$\si$--model requires that the target manifold is K\"ahler, that
the mirror
vacuum cannot even be $N=1$ spacetime supersymmetric.
It follows that the class of Calabi--Yau manifolds is not the
appropriate setting by a long shot in which to discuss mirror symmetry
and
the question arises  what the proper framework might be.

In this review I discuss recent work \cite{rs5} which
shows the existence of a new class of manifolds
which generalizes the class of Calabi--Yau spaces of complex dimension
$D_{crit}$ in a natural way. The manifolds involved are of complex
dimension $(2k+D_{crit})$ and have a positive first Chern class which
is quantized in
multiples of the degree of the manifold. Thus they do not describe,
a priori, consistent string groundstates. Surprisingly however, it is
possible to derive from these higher dimensional
manifolds the spectrum of  critical string vacua. This
can be done not only for the generations but also for the
antigenerations. For particular types of
these new manifolds it is also possible to construct the corresponding
$D_{crit}$--dimensional Calabi--Yau manifold directly from the
$(2k+D_{crit})$--dimensional space.

This new  class of manifolds is, however, not in one to one
correspondence
with the class of Calabi--Yau manifolds as it also contains manifolds
which describe string vacua
that do not contain massless modes corresponding to antigenerations.
It is precisely this new type of manifold that is
needed in order to construct mirrors of rigid Calabi--Yau manifolds
 without generations.
\vglue 1cm
{\bf\noindent 2. Higher Dimensional Manifolds with Quantized
Positive First Chern Class }
\vglue 0.4cm
\noindent
Consider the class of manifolds of complex dimension $N$ embedded in a
weighted projective space $\IP_{(k_1,\dots, k_{N+2})}$ as hypersurfaces
\beq
M_{N,d}~ \equiv ~ \{p(z_1,\dots,z_{N+2})=0\}~\cap ~
\IP_{(k_1,\dots, k_{N+2})}            \nn
\eeq
defined as the zero set of some transverse polynomial $p$ of degree $d$.
Here the $k_i$ are the weights of the ambient weighted projective space.
The set of hypersurfaces determined by such polynomials will be
denoted by
\beq
\IP_{(k_1,k_2,\dots \dots ,k_{N+2})}[d]      \nn
\lleq{phyp}
and called a configuration.
Assume that for the hypersurfaces (\ref{phyp}) the weights $k_i$ and
the degree $d$ are related via the constraint
\beq
\sum_{i=1}^{N+2} k_i =  Qd,
\lleq{pquant}
where $Q$ is a positive integer. Relation (\ref{pquant}) is the defining
property of the class of spaces to be considered below. It is a rather
restrictive condition in that it excludes many types of varieties which
are transverse and even smooth but are not of physical relevance
\fnote{2}{It will become clear below that this definition is rather
          natural in the context of the theory of Landau--Ginzburg
          string vacua with an arbitrary number of scaling fields. A
         particular simple manifold in this class, the cubic sevenfold
       $\IP_8[3]$, has been the subject of recent investigations
       \cite{philbalt}\cite{pel}\cite{cumrun2}.}.
A simple example is the Fermat hypersurface
\beq
\IP_{(420,280,210,168,140,120,105)}[840] \ni
\{p= \sum_{i=1}^7 z_i^{i+1} =0 \}
\lleq{excluded}
which is a rather nice, transverse, i.e. quasismooth manifold. It is
also interesting from a different point of view. A curious aspect of
Calabi--Yau hypersurfaces is that they are automatically what is called
{\it well formed}, i.e. they do not contain orbifold singularities that
are surfaces (in the case of threefolds). More generally this fact
translates into the statement that the only resolutions that have to
be performed are so--called small resolutions, i.e. the singular set
are of codimension larger than one. The same is true for the
higher dimensional manifolds defined above whereas the manifold
(\ref{excluded})  contains the
singular 4--fold $S=\IP_{(210,140,105,84,70,60)}[420]$.

Alternatively, manifolds of the type above may be characterized
via a curvature constraint. Because of (\ref{pquant}) the first
Chern class is given by
\beq
c_1(M_{N,d}) =(Q-1)~c_1(\cN)
\lleq{c1quant}
where $c_1(\cN)=dh$ is the first Chern class of the normal bundle
$\cN$ of
the hypersurface $M_{N,d}$ and $h$ is the pullback of the K\"ahler form
${\rm H}\in {\rm  H}^{(1,1)}\left(\IP_{(k_1,\dots,k_{N+2})}\right)$
of the
ambient space. Hence the first Chern class is quantized in multiples
of the
degree of the hypersurface $M_{N,d}$.
For $Q=1$ the first Chern class vanishes and the manifolds
for which (\ref{pquant}) holds are Calabi--Yau manifolds, defining
consistent groundstates of the supersymmetric closed string.
For $Q > 1$ the first Chern class is nonvanishing and therefore these
manifolds cannot possibly describe vacua of the critical string, or
so it seems.

It turns out that these spaces are closely
related to string vacua of complex critical dimension
\beq
D_{crit} = N-2(Q-1)
\lleq{critdim}
i.e. the critical dimension is offset by twice the coefficient of the
first Chern class of the normal bundle of the hypersurface.
The evidence for this is twofold. First it is possible to
derive from these higher dimensional manifolds the massless spectrum
of critical vacua. Furthermore it is shown that for certain subclasses
of hypersurfaces of type (\ref{pquant}) it is possible to construct
Calabi--Yau manifolds $M_{CY}$ of dimension $D_{crit}$  and complex
codimension
\beq
codim_{\IC} (M_{CY}) =Q      \nn
\eeq
directly from these manifolds. In terms of the critical dimension and
the codimension the class of manifolds to be investigated below can be
described as the projective configurations
\beq
\IP_{(k_1,\dots ,k_{(D_{crit}+2Q)})}
\left[\frac{1}{Q}\sum_{i=1}^{D_{crit}+2Q} k_i\right] .
\lleq{newmfs}

As mentioned already in the introduction the class of spaces defined by
(\ref{newmfs}) contains manifolds with no antigenerations
and hence it is necessary to have some way other than Calabi--Yau
manifolds
to represent string groundstates if one wants to compare them with the
higher
dimensional manifolds. One possible way to do this is to relate them to
Landau--Ginzburg theories: any manifold of type (\ref{phyp}) can be
viewed
as a projectivization via a weighted equivalence defined on an affine
noncompact hypersurface  defined by the same polynomial
\beq
\IC_{(k_1,...,k_{N+2})}\left[d\right] \ni \{p(z_1,...,z_{N+2})=0\}.
\lleq{affvar}
Because the polynomial $p$ is assumed to be transverse in the projective
ambient
space the affine variety has a very mild singularity: it has an isolated
singularity at the origin defining what is called a catastrophe in the
mathematics literature.

The complex variables $z_i$ parametrizing the ambient space are to be
viewed as the field theoretic limit
$\varphi_i(z,\bz) = z_i$ of the lowest components of the order
parameters
$\Phi_i(z_i,\bz_i,\th^{\pm}_i,\bth^{\pm}_i)$,
described by chiral $N=2$ superfields of a 2--dimensional
Landau--Ginzburg theory defined by the action
\beq
\int d^2z d^2\th d^2\bth~ K(\Phi_i,\bPhi_i) +
\int d^2z d^2\th~ W(\Phi_i) + c.c.                      \nn
\eeq
where $K$ is the K\"ahler potential and $W$ is the superpotential.
It was the important insight of Martinec \cite{m} and Vafa and
Warner \cite{vw}
that such Landau--Ginzburg theories are useful for the understanding of
 string
vacua and also that much information about such groundstates is already
encoded in the associated catastrophe (\ref{affvar}). A crucial
piece of
information about a vacuum, e.g., is its central charge. Using a result
from singularity theory, it is easy to derive that the central charge
of the conformal fixed point of the LG theory is
\beq
c=3\sum_{i=1}^{N+2} \left(1-2q_i\right),
\eeq
where $q_i =k_i/d$ are the U(1) charges of the superfields. It is
furthermore  possible
to derive the massless spectrum of the GSO projected fixed of the
LG theory,
defining
the string vacuum, directly from the catastrophe (\ref{affvar}) via a
procedure described by Vafa \cite{cumrun1}.

The manifolds (\ref{newmfs}) therefore correspond to LG theories of
central charge
\beq
c=3(N-2(Q-1))=3D_{crit}
\eeq
where the relation (\ref{critdim}) has been used.

In certain benign situations the subring of monomials
of charge 1 in the chiral ring describes the generations of the
vacuum \cite{philip1}. For this to hold at all it is important that
the GSO projection
is the canonical one with respect to the cyclic group $\ZZ_d$,
the  order of which is the degree $d$ of the superpotential
\fnote{3}{It does not hold for projections that involve orbifolds
with respect
       to different groups such as those discussed in \cite{bgh}.
This is
       to be expected as these modified projections
       can be understood as orbifolds of canonically constructed
       vacua. The additional moddings generate singularities the
resolution
       of which introduces, in general,
  additonal modes in both sectors, generations and antigenerations.}.
Thus the generations are easily derived for this subclass of
 theories in (\ref{newmfs})  because the polynomial ring is identical
to the chiral ring of the corresponding Landau--Ginzburg theory.
In general a more sophisticated analysis, involving the resolution of
higher dimensional singularities, will have to be done \cite{rstobe}.

It remains to extract the second cohomology. In a Calabi--Yau manifold
there are no holomorphic  2--forms and hence all of the second
cohomology
is in $H^{(1,1)}$. Because of Kodaira's vanishing theorem the same
is true
for manifolds with positive  first Chern class and therefore for the
manifolds under discussion. At first sight it might appear hopeless
to find a construction corresponding to the analysis of (2,1)--forms
because of the following example which involves the orbifold of a
3--torus.

Consider the
orbifold $T_1^3/\ZZ_3^2$ where the two actions are defined as
$(z_1,z_4) \longrightarrow  (\a z_1,\a^2 z_4)$, all other coordinates
invariant
and $(z_1,z_7) \longrightarrow (\a z_1,\a^2 z_7) $, all other invariant.
Here $\a$ is the third root of unity.
The resolution of the singular orbifold leads to a Calabi--Yau manifold
with
 84 antigenerations and no generations \cite{gvw}.
This is precisely the mirror flipped spectrum of the exactly solvable
tensor
model $1^9$ of 9 copies of $N=2$ superconformal minimal models at level
$k=1$ \cite{doron1} which can be described in terms of the
Landau--Ginzburg
potential $W=\sum z_i^3$ which belongs to the configuration
$\IC_{(1,1,1,1,1,1,1,1,1)}[3]$. After imposing the GSO projection by
modding out a $\ZZ_3$ symmetry this Landau--Ginzburg theory leads to the
same spectrum as the $1^9$ theory.

This Landau--Ginzburg theory clearly is a mirror candidate for the
resolved torus orbifold just mentioned
\cite{philbalt}\cite{pel}\cite{cumrun2}
and the question arises whether a manifold corresponding to
this LG potential can be found. Since the theory does not contain modes
corresponding to (1,1)--forms it appears that the manifold cannot be
K\"ahler and hence not projective. Thus it appears that the
7--dimensional
 manifold $\IP_8[3]$ whose polynomial ring is  identical to the
chiral ring
of the LG theory is merely useful as an auxiliary device
in order to describe one sector of the critical LG string vacuum.
Even though there exists a precise identity between the Hodge numbers
in the
middle cohomology group of the higher dimensional manifold and the
middle
dimension of the cohomology of the Calabi--Yau manifold this is not
the case for the second cohomology group.
\vglue 1cm
\noindent
{\bf\noindent 3. Noncritical Manifolds and Critical Vacua}
\vglue 0.4cm
\noindent
It turns out however, that by looking at the manifolds
(\ref{newmfs})  in a
slightly different way it is nevertheless possible to extract the
second
cohomology in a canonical manner (even if there is {\it none}).
The way this works is as follows: the manifolds of type
(\ref{newmfs}) will, in general, not be described by smooth spaces but
will have singularities which arise from the projective identification.
The basic idea now is to associate the existence of antigenerations in
a {\it critical}
string vacuum with the existence of singularities in these higher
dimensional {\it noncritical} spaces.

Because the structure of these geometrical
singularities depends on the
precise form of the polynomial constraint it is difficult to
prove the correctness of this idea in full generality.
vacua
Instead I will, in the following, make this idea more precise and
illustrate  how it works with a few particularly simple classes of
theories, leaving a more detailed investigation of other types of
manifolds
for a more extensive discussion \cite{rstobe}. As an
unexpected bonus this derivation will provide new insight
into the Landau--Ginzburg/Calabi--Yau connection.

Consider again the simple example related to the tensor
model $1^9$. Its LG theory is described by $\IC^*_9[3]$ the naive
compactification of which leads to
\beq
\IP_8[3]\ni \{\sum_{i=1}^9 z_i^3=0\}.   \nn
\lleq{ex1}
Counting monomials leads to the spectrum of 84 generations found
previously
for the corresponding string vacuum and because this manifold is
smooth {\it no} antigenerations are expected in this model!
Hence there does not exist a Calabi--Yau manifold that describes the
groundstate
\fnote{4}{ It would seem that a  generalization of
           this 7--dimensional smooth manifold is the infinite class of
          models
           $\IC_{(1,1,1,1,1,1,1,1,1+3q)}[3+q]$, but since the
           manifolds (\ref{newmfs}) are required to be transverse the
only
           possibility is $q=0$.}
{}.
A second theory in the space of all LG vacua with no antigenerations is
\beq
(2^6)_{A^6}^{(0,90)} \equiv \IC^*_{(1,1,1,1,1,1,2)}[4] \ni
\{\sum_{i=1}^6 z_i^4 + z_7^2 =0\}                      \nn
\lleq{ex2}
with an obviously smooth manifold $\IP_{(1,1,1,1,1,1,2)}[4] $.

Vacua without antigenerations are rather exceptional however; the generic
groundstate will have both sectors, generations and antigenerations.
The idea described above to derive the antigenerations works for
higher dimensional manifolds corresponding to different types of critical
vacua but in the following we will illustrate it with two types of such
manifolds. A more detailed analysis can be found in \cite{rstobe}.

To be concrete consider the
exactly solvable tensor theory $(1\cdot 16^3)_{A_2\otimes E_7^3}$ with
35 generations and 8 antigenerations which
corresponds to a Landau--Ginzburg theory belonging to the configuration
\beq
\IC^*_{(2,3,2,3,2,3,3)}[9]^{(8,35)}
\eeq
and which induces, via projectivization, a 5--dimensional weighted
hypersurface
\beq
\IP_{(2,2,2,3,3,3,3)}[9] \ni
\{p=\sum_{i=1}^3 (y_i^3x_i+x_i^3)+x_4^3=0\}.
\lleq{ex3}
with orbifold singularities
\bea
\ZZ_3 &:& \IP_3[3] \ni \{p_1=\sum_{i=1}^4 x_i^3=0\} \nn \\
\ZZ_2 &:& \IP_2.
\eea
The $\ZZ_3$--singular set is a smooth cubic surface which supports
 seven (1,1)--form as can be easily shown. The
$\ZZ_2$ singular set is just the projective plane and therefore
adds one further (1,1)--form.  Hence the singularities induced on the
hypersurface by the singularities of the ambient weighted projective
space
give rise to a total of eight  (1,1)--forms. A simple count leads to the
result that the subring of monomials of charge 1 is of dimension 35.
Thus we have derived the
spectrum of the critical theory from the noncritical manifold
(\ref{ex3}).

It is presumably possible to derive this result via a surgery process
on the singular space (\ref{ex3}) but more important is, at this point,
 that the idea introduced above of relating the spectrum of the string
vacuum to the singularity structure of the noncritical manifold also
makes it possible to derive from these higher dimensional
manifolds the Calabi--Yau manifold of critical dimension! Thus a
canonical
prescription is obtained which also allows to pass from the
Landau--Ginzburg theory to its geometrical counterpart.

This works as follows:
 Recall that the structure of the singularities of the weighted
hypersurface
just involved part of the superpotential, namely the cubic polynomial
$p_1$ which determined the $\ZZ_3$ singular set described by a surface.
The superpotential thus splits naturally into the two parts
\beq
p=p_1 + p_2
\eeq
where $p_2$ is the remaining part of the polynomial. The idea now is to
consider the product $\IP_3[3] \times \IP_2$ where the factors are
determined by
the singular sets of the higher dimensional space and to impose on this
4--dimensional space a constraint described by  the remaining part of
the polynomial which did not
take part in constraining the singularities of this ambient space.
In the
case at hand this leaves a polynomial of bidegree $(3,1)$ and hence
we are
lead to a manifold embedded in
\beq
\matrix{\IP_2 \hfill \cr \IP_3\cr}
\left[\matrix{3&0\cr 1&3\cr}\right]
\lleq{ex3mine}
defined by polynomials
\bea
p_1 &=& y_1^3x_1 + y_2^3x_2 +y_3^3x_3 \nn \\
p_2 &=& \sum_{i=1}^4 x_i^3
\eea
which is precisely the manifold constructed in \cite{rs1}, the exactly
solvable
model of which was later found in \cite{doron2}. Thus we have found how
to construct
from the noncritical manifold (\ref{ex3}) the critical Calabi--Yau
manifold.

A subclass of manifolds of a different type which can be discussed in
this framework rather naturally is defined by
\beq
\IP_{(2k,K-k,2k,K-k,2k_3,2k_4,2k_5)}[2K]
\lleq{niceclass}
where $K=k+k_3+k_4+k_5$ and it is assumed, for simplicity,
 that  $K/k$ and $K/k_i$ are integers. The  potentials are
\beq
W=\sum_{i=1}^2(x_i^{K/k}+x_iy_i^2) +x_3^{K/k_3}+x_4^{K/k_4} +x_5^{K/k_5}.
\eeq
The singularities in these manifolds are of two types,
\bea
\ZZ_2 &:&~~\IP_{(k,k,k_3,k_4,k_5)}[K] \nn \\
\ZZ_{K-k} &:&~~ \IP_1.
\eea
where the constraint of degree $K$ is given by
\beq
p_1 = \sum_{i=1}^5 x^{K/k_i}.
\eeq
The $\ZZ_2$--singular set is 3--fold with positive first Chern class
embedded in weighted $\IP_4$ whereas the $\ZZ_{K-k}$ singular set is
just the sphere $S^2 \sim \IP_1$.

To construct the corresponding critical manifolds note
that the structure of the singularities of the weighted hypersurface
just involved part of the superpotential, namely the quartic polynomial
$p_1$ which determined the $\ZZ_2$ singularity set described by a
3--fold.
The superpotential thus splits naturally into the two parts
$p=p_1 + p_2$
where $p_2$ is the remaining part of the polynomial. The idea now is to
consider again
the product $\IP_{(k,k,k_3,k_4,k_5)}[K] \times \IP_1$
of  singular sets of the higher dimensional space and to impose on this
4--dimensional space a constraint described, as before, by the remaining
part $p_2$ of the polynomial which did not
take part in constraining the singularities of this ambient space.
In the
case at hand this leaves a polynomial of bidegree $(k,2)$ and hence
we are
lead to a manifold embedded in
\beq
\matrix{\IP_1 \hfill \cr \IP_{(k,k,k_3,k_4,k_5)}\cr}
\left[\matrix{2&0\cr k&K\cr}\right]
\lleq{s3class}
defined by polynomials
\bea
p_1 &=& y_1^2x_1 + y_2^2x_2 \nn \\
p_2 &=& x_1^{K/k}+ x_2^{K/k} + x_3^{K/k_3} + x_4^{K/k_4} + x_5^{K/k_5}.
\eea
That this correspondence is in fact  correct  can be inferred from the
work of \cite{rs3}
where it was shown that these codimension--2 weighted CICYs correspond
to $N=2$ minimal exactly solvable tensor models  of the type
\beq
\left[ 2\left(\frac{K}{k}-1\right)\right]_D^2 \cdot
\prod_{i=3}^5  \left(\frac{K}{k_i}-2\right)_{A}.
\eeq
where the subscripts indicate the affine invariants chosen for the
individual levels.

The general picture that emerges from these constructions then is the
following: embedded in the higher dimensional manifold is a submanifold
which is fibered where the base and the fibres are determined by the
singular sets of the ambient manifold. The Calabi--Yau manifold itself
is a hypersurface embedded in this fibered  submanifold. A heuristic
sketch of the geometry is shown in the Figure 1.


The examples above illustrate the simplest situation that can appear.
In more complicated manifolds the singularity structure will consist
of hypersurfaces whose fibers and/or base themselves are fibered,
leading to
an iterative procedure. The submanifold to be considered will, in
those
cases, be of codimension larger than one and the Calabi--Yau
manifold will be described by a submanifold with codimension
larger than one as well. To illustrate this point consider the 7--fold
\beq
\IP_{(1,1,6,6,2,2,2,2,2)}[8]~\ni~
\left\{\sum_{i=1}^2 \left(x_i^2y_i + y_iz_i +z_i^4\right)
 +z_3^4+z_4^4+z_5^4=0\right\}
\eeq
which leads to the $\ZZ_2$ fibering
$\IP_1\times \IP_{(3,3,1,1,1,1,1)}[4]$ which in turn leads to the
$\ZZ_3$ fibering $\IP_1\times \IP_1 \times \IP_4[4]$. Following the
splits of the potential thus leads to the Calabi--Yau configuration
\beq
\matrix{\IP_1\cr \IP_1 \cr \IP_4\cr}
\left[\matrix{2&0&0\cr
              1&1&0\cr
              0&1&4\cr}\right] \ni
\left\{ \begin{array}{c l}
        p_1=& \sum_{i=1}^2 x_i^2y_i=0 \\
        p_2=& \sum_{i=1}^2 y_iz_i =0 \\
        p_3=& \sum_{j=1}^5 z_i^4 =0
         \end{array}
\right\}
\lleq{codim3}
which is of codimension 3. This example also shows that there are
nontrivial relations between these higher dimensional manifolds.
The way to see this is via the process of splitting and contraction
of Calabi--Yau manifolds introduced in ref. \cite{cdls}.
It can be shown in fact that the Calabi--Yau manifold (\ref{codim3})
is an ineffective split of a Calabi--Yau manifold in the class
(\ref{niceclass}). Thus there also exists a corresponding
relation between the higher dimensional manifolds.
\vglue 1cm
{\bf\noindent 4. Generalization to Arbitrary Critical Dimensions}
\vglue 0.4cm
\noindent
Even though the examples discussed in the previous section were all
concerned with 6--dimensional Calabi--Yau manifolds and the way they
are embedded in the new class of spaces, it should be clear that
the ideas presented are not specific to this dimension.
 Instead of considering compactifications down to the
physical dimension, 4, one might contemplate compactifying down to
2, 6 or
8 dimensions, or else, discuss the class of manifolds above not in the
context of compactification at all.

To illustrate this point consider the infinite class of manifolds
\beq
\IP_{(2,n-1,2,n-1,2,\dots ,2)}[2n]
\eeq
of complex dimension $n+1$, defined by polynomials
\beq
p=\sum_{i=1}^2(x_i^n+x_iy_i^2) + x_3^n + \cdots + x_{n+1}^n.
\lleq{infser}
According to the considerations above these spaces are related to
Calabi--Yau manifolds embedded in
\beq
\matrix{\IP_1\cr \IP_n\cr}\left[\matrix{2&0\cr 1&n\cr}\right]
\eeq
via the equations
\bea
p_1 &=& y_1^2x_1+y_2^2x_2 \nn \\
p_2 &=& \sum_{i=1}^{n+1} x_i^n.
\eea

The simplest example is, of course, the case $n=2$ where the higher
dimensional manifold is a 3--fold described by
\beq
\IP_{(2,1,2,1,2)}[4]~\ni ~ \{\sum_{i=1}^2 (z_i^2+z_iy_i^2)+z_3^2=0\}
\eeq
with a $\ZZ_2$ singular set isomorphic to the sphere
$\IP_2[2]\sim \IP_1$
which contributes one (1,1)--form, the remaining one being provided
by the
$\IP_1$ defined by the remaining coordinates.
The singularity structure of the 3--fold then relates this space to the
complex torus described by the algebraic curve
\beq
\matrix{\IP_1 \cr  \IP_2}\left[\matrix{2 &0\cr  1 & 2\cr}\right].
\eeq

It should be remarked that the Landau--Ginzburg theory corresponding
to this
theory derives from an exactly solvable tensor model $(2^2)_{D^2}$
described by two $N=2$ superconformal minimal theories at level $k=2$
equipped with the affine D--invariant.

It is of interest to consider the cohomology groups of the 3--fold
itself.
With the third Chern class $c_3=2h^3$ the Euler number of the
singular space is
\beq
\chi_s = \int c_3 =1
\eeq
and hence the Euler number of the resolved manifold is
\beq
\tchi=1-(2/2)+2\cdot  2 =4.
\eeq
Since the singular set is a sphere its resolution contributes just one
(1,1)--form and hence the second Betti number becomes $b_2=2$.
With  $\tchi=2(1+h^{(1,1)})-2h^{(2,1)}$ it follows that
\beq
h^{(2,1)}=1.
\eeq

The case $n=3$ is particularly illuminating because it involves a higher
dimensional manifold that is smooth
\beq
\IP_5[3]
\eeq
and hence it is easy to determine the
cohomology  groups of this space and to  compare it with the spectrum
of K3
\beq
K3=\matrix{\IP_1\cr \IP_3\cr} \left[\matrix{2&0\cr 1&3\cr}\right],
\eeq
which consists of $h^{(0,0)}=h^{(2,2)}=h^{(2,0)}=h^{(0,2)}=1$ and
$h^{(1,1)}=20$, all other Hodge numbers are zero. Hence the Euler number
becomes $\chi(K3)=24$.

 The Euler  number for the noncritical manifold
is easily computed to be $\chi=27$. Since the manifold is K\"ahler
$h^{(p,q)}=h^{(q,p)}$ and because of Poincar\'e duality $b_p = b_{8-p}$.
Because the manifold has positive first Chern  class it follows from
Kodaira's vanishing theorem that $h^{(p,0)}=0$ for $p\neq 0$ and via
Lefshetz' hyperplane theorem it is known that below the middle dimension
all the cohomology is inherited from the ambient space and therefore
the  only  nonvanishing cohomology groups lead to
$h^{(0,0)}=h^{(1,1)}=1$.
It can be shown that $h^{(3,1)}=h^{(1,3)}=1$
\fnote{5}{I'm grateful to P.Candelas and T.H\"ubsch for explanations
regarding
         this computation.}
 and therefore the only remaining cohomology is in $H^{(2,2)}$. Since
\beq
\chi = 2(b_0+b_2) + b_4 = 6+h^{(2,2)}
\eeq
it follows that $h^{(2,2)}=21=20+1$. Thus we have obtained the spectrum
of K3 plus one additional mode which always appears in this construction.

This example is also useful because it indicates a generalization of the
considerations of the previous section. The  surprising new feature of
this
manifold is that even though the higher dimensional manifold did not
have
any orbifold singularities it was nevertheless possible to split it in
such a way as to construct a Calabi--Yau manifold from it. This was
possible
because the defining equation was not of Fermat type but involved
couplings
between the fields. Because of this the manifold featured a new $\ZZ_2$
symmetry not present in the Fermat hypersurface and it is this new
symmetry that dictated   how to perform the split. This indicates that
even for smooth higher dimensional manifolds it is possible to relate
them
to Calabi--Yau manifolds  once one moves away from the symmetric point.
Table 1 containes a few other examples of how to relate different
singular 4--folds to K3 representations.

\vskip .3truein

{\scriptsize
\begin{center}
\begin{tabular}{|| l l l l||}
\hline
ECFT  &Projective Manifold   &Singularities   &CY \tabroom \\
\hline
\hline
$(1^6)_{D^2\otimes A^4}$
&$\IP_5[3]$    &$\ZZ_2:  \IP_1$  & \tabroom \\
&$\sum (z_i^3+z_iy_i^2)+z_3^3+z_4^3$
                             &
&$\matrix{\IP_1 \cr  \IP_3}\left[\matrix{2 &0\cr  1 & 3\cr}\right]$ \\
[3ex]
$(6^2\cdot 2)_{D^2\otimes A}$
&$\IP_{(2,3,2,3,2,4)}[8]$    &$\ZZ_2:  \IP_{(1,1,1,2)}[4]$ &\tabroom \\
&$\sum (z_i^4+z_iy_i^2)+z_3^4+z_4^2$
                             & $\ZZ_3: \IP_1 $
  &$\matrix{\IP_1\hfill\cr  \IP_{(1,1,1,2)}}
                  \left[\matrix{2 &0\cr  1 & 4\cr}\right]$ \\ [3ex]
$(10^2\cdot 1)_{D^2\otimes A}$
&$\IP_{(2,5,2,5,4,6)}[12]$
                             &$\ZZ_2:  \IP_{(1,1,2,3)}[6]$  & \tabroom \\
&$\sum (z_i^6+z_iy_i^2)+z_3^3+z_4^2$
                             &$\ZZ_5: \IP_1 $
&$\matrix{\IP_1\hfill  \cr \IP_{(1,1,2,3)} \cr}
\left[\matrix{2 &0\cr  1 & 6\cr}\right]$ \\
\hline
\hline
\end{tabular}
\end{center}
}
\noindent
{\bf Table 1:} {\it Examples of dimension $D_{\IC}=2$}

\vskip .1truein
An example involving a 4--dimensional critical manifold of a different
 type is defined by the polynomial
\beq
p=\sum_{i=1}^3 \left(x_i^3+x_iy_i^3\right) +
  \sum_{j=4}^5 x_i^6
\eeq
 which corresponds to the tensor model
$(16^3 \cdot 4)_{E^3 \otimes D^2}$ with central charge $c=12$ and
belongs to the configuration
\beq
\IP_{(6,4,6,4,6,4,3,3)}[18].
\eeq
The critical manifold derived from this 6--fold belongs to
the configuration class
\beq
\matrix{\IP_2 \hfill\cr \IP_{(2,2,2,1,1)}\cr}
\left[\matrix{3&0\cr 2&6\cr}\right]
\eeq
which is indeed a Calabi--Yau deformation class.

Again it should be emphasized that the construction is not restricted
to the
infinite series defined in (\ref{infser}) as the final example
illustrates.
A five--dimensional critical vacuum is obtained by considering the
Landau--Ginzburg potential
\beq
W= \sum_{j=1}^2 \left(u_i^3 + u_iv_i^2\right) +
    \sum_{i=3}^5 \left(u_i^3 + u_iw_i^3\right)
\eeq
which corresponds to the exactly solvable model
$(16^3\cdot 4^2)_{E_7^3 \otimes D^2}$. The nine--dimensional noncritical
manifold
\beq
\IP_{(3,2,3,2,3,2,3,3,3,3)}[9]
\eeq
leads, via its singularity structure, to the five--dimensional critical
manifold
\beq
\matrix{\IP_1\hfill  \cr \IP_2 \cr \IP_4\cr}
\left[\matrix{2 &0 &0\cr  0 & 3 &0\cr 1 & 1 & 3\cr}\right] .
\eeq
It is again crucial that a non--Fermat polynomial was chosen for the
last four coordinates in the noncritical manifold.
\vglue 1cm
{\bf\noindent 5. Conclusion \hfil}
\vglue 0.4cm
\noindent
Mirror symmetry cannot be understood in the framework of Calabi--Yau
manifolds.
Thus, beyond the class of such spaces, there must exist a space of a
new
type of noncritical manifolds which contain information about critical
vacua,
such as the mirrors of rigid Calabi--Yau manifolds. Mirrors of spaces
with
both sectors, antigeneration and generations, are again of Calabi--Yau
type
and hence the noncritical manifolds which correspond to such
groundstates
should make contact with Calabi--Yau manifolds in some manner.

What has been  shown in \cite{rs5} is that the class (\ref{newmfs}) of
higher
dimensional K\"ahler
manifolds with positive first Chern class, quantized in a particular way,
generalizes the framework of Calabi--Yau vacua in the desired way: For
particular types of such noncritical manifolds Calabi--Yau manifolds of
critical dimension are embedded algebraically in a fibered submanifold.
For string vacua which cannot be described by
K\"ahler manifolds and which are mirror candidates of rigid Calabi--Yau
manifolds the higher dimensional manifolds still lead to the
spectrum of the critical vacuum and a rationale emerges that explains
why a
Calabi--Yau representation is not possible in such theories.
Thus these manifolds of dimension $c/3 +2k$
 define an appropriate framework in which to discuss mirror symmetry.

There are a number of important consequences that follow from the
results
of the previous sections. First it should be realized that the relevance
of noncritical manifolds suggests the generalization of a conjecture
regarding
the relation between (2,2) superconformal field theories of central
charge
$c=3D$, $D\in \IN$, with N=1 spacetime supersymmetry on the one hand
and K\"ahler manifolds
of complex dimension $D$ with vanishing first Chern class on the other.
It was suggested
by Gepner that this relation is 1--1. It follows from the results
above that
instead superconformal theories of the above type are in correspondence
with K\"ahler manifolds of dimension $c/3 +2k$ with a first Chern
class quantized in multiples of the degree.

A second consequence is that the ideas of section 3 lead, for a large
class of Landau--Ginzburg theories, to a new canonical prescription
for the
construction of the critical manifold, if it exists, directly from
the 2D
field theory.

Recently Batyrev \cite{vitja} introduced a new construction of
mirrors of
Calabi--Yau manifolds based on dual polyhedra. His method appears to
apply only to manifolds defined by one polynomial in a weighted
projective
space or products thereof. The method of toric geometry that is used in
\cite{vitja} is however not restricted to Calabi--Yau manifolds
\cite{vitja2} and therefore the
constructions described in sections 3 and 4 lead to the exciting
possibility
of extending Batyrev's results to Calabi--Yau manifolds of codimension
larger
than one by proceeding via noncritical manifolds.

A final remark is that in this framework the role played by the
dimension
of the manifolds  becomes of secondary importance. This is as it
should be,
at least for an effective theory, which tests only matter content and
couplings. It is, however, somewhat mysterious that via ineffective
splittings manifolds of different
dimension describe one and the same critical vacuum.

It is clear that the emergence in string theory of manifolds with
quantized first Chern class should be understood better. The results
described here are a first step in this direction. They indicate
that these manifolds are not just  auxiliary devices but may be as
physical as  Calabi--Yau manifolds of critical dimension.
In order to probe the structure of these models in more depth it is
important to get further insight into the complete spectrum of
these theories and to compute the Yukawa couplings of the fields.
It is clear from the results presented here that the spectra of the
higher dimensional manifolds contain additional modes beyond those that
are related to the generations and antigenerations of the critical
vacuum
and the question arises what physical interpretation these fields have.

A better grasp on the complete spectrum of these spaces should also
give insight into a different, if not completely independent, approach
toward a deeper understanding of these higher dimensional manifold,
which
is to attempt the construction of consistent $\si$--models defined via
these spaces. Control of the complete spectrum will shed light
on the precise relation between the $\si$--models associated to the
noncritical manifolds and critical $\si$--models.
\vglue 1cm
{\bf \noindent 5. Acknowledgements \hfil}
\vglue 0.4cm
I'm grateful to CERN for hospitality and the theoretical theorists there
for lively discussions, in particular Per Berglund, Philip Candelas,
Wolfgang Lerche and Jan Louis. It is a pleasure to thank Herbert Clemens,
Tristan H\"ubsch, Cumrun Vafa and Nick Warner for conversations and
R.Hartshorne for asking the right question. I'm also grateful to the
Aspen Center of Physics for hospitality.
\vglue 1cm
{\bf\noindent 6. References \hfil}
\vglue 0.4cm

\end{document}